# Magnetic field dependence of the nonlinear magnetic response and tricritical point in the monoaxial chiral helimagnet $Cr_{1/3}NbS_2$

E. M. Clements[1], Raja Das[1], Manh-Huong Phan[1,*], Ling Li[2], Veerle Keppens[2], David Mandrus[2], Michael Osofsky[3], and Hariharan Srikanth[1,*]

[1]Department of Physics, University of South Florida, Tampa, FL 33620, USA
[2]Department of Materials Science and Engineering, University of Tennessee, Knoxville, Tennessee 37996, USA
[3]Naval Research Laboratory, Washington, D.C. 20375, USA

## Abstract

We present a comprehensive study of the magnetization dynamics and phase evolution in the chiral helimagnet $Cr_{1/3}NbS_2$, which realizes a chiral soliton lattice (CSL). The magnetic field dependence of the ac magnetic response is analyzed for the first five harmonic components, $M_{n\omega}(H)$ ($n = 1 - 5$), using a phase sensitive measurement over a frequency range, $f = 11 - 10,000$ Hz. At a critical field, the modulated CSL continuously evolves from a helicity-rich to a ferromagnetic domain-rich structure, where the crossover is revealed by the onset of an anomalous nonlinear magnetic response that coincides with extremely slow dynamics. The behavior is indicative of the formation of a spatially coherent array of large ferromagnetic domains which relax on macroscopic time-scales. The frequency dependence of the ac magnetic loss displays an asymmetric distribution of relaxation times across the highly nonlinear CSL regime, which shift to shorter time-scales with increasing temperature. We experimentally resolve the tricritical point at $T_{TCP}$ in a temperature regime above the ferromagnetic Curie temperature which separates the linear and nonlinear magnetic regimes of the CSL at the phase transition. A comprehensive phase diagram is constructed which summarizes the features of the field and temperature dependence of the magnetic crossovers and phase transitions in $Cr_{1/3}NbS_2$.

## I. INTRODUCTION

The spatially-modulated magnetic states that arise in noncentrosymmetric magnetic materials with strong spin-orbit coupling have gained significant interest due to the stability and high degree of tunability of their symmetry-protected spin textures. [1–6] In $Cr_{1/3}NbS_2$, the chiral helimagnetic (CHM) structure, with period $L(0) = 48$ nm, propagates over a remarkably large spatial range. This is due to the close coupling of the localized magnetic moments with the underlying crystal lattice via the Dzyaloshinskii-Moriya interaction [7,8] and the high uniaxial anisotropy which fixes the spin helix along the crystallographic $c$ axis. [9–11] The layered structure of $Cr_{1/3}NbS_2$ consists of $2H$-type planar $NbS_2$ with Cr atoms intercalated between planes and belongs to the hexagonal space group $P6_322$, which lacks inversion symmetry. [12] The localized moments of the $Cr^{3+}$ ions ($S = 3/2$) are oriented in the $ab$ plane and exhibit strong single-ion anisotropy. [13,14] Under a magnetic field applied perpendicularly to the chiral axis, the harmonic spiral structure continuously crosses over into a nonlinear chiral soliton lattice (CSL), as illustrated in Fig. 1(a). [15,16] According to a quasi-1D model, [6,17] the CSL can be described as a periodic chain of ferromagnetic domains separated by 360° domain walls, called solitons. [15,18] The physical realization of the CSL in $Cr_{1/3}NbS_2$ was first observed by Togawa and coworkers via a Lorentz microscopy experiment. [16] In recent years, a number of studies have uncovered fundamental properties of the symmetry-broken magnetic state. The sliding motion of the highly coherent CSL structure in the presence of an ac magnetic field has been theoretically shown to amplify the physically observable spin motive force, which is in direct proportion to the number of solitons along the spin chain. [19,20] Several studies which followed have shown the effects of confinement on both its collective dynamics [21] and of the topologically protected CSL, in which the number of solitons can be discretely controlled. [22–24] Thus, as a candidate for spintronics applications, a clear understanding of the magnetic field dependence of the ac magnetic response of the CSL is essential in controlling spins and spin transport for magnetoelectronic devices.

The chiral magnetic phase is characterized by a robust spin coherence in which both the amplitude and phase of the order parameter display long-range order. [17] In the presence of a magnetic field, the modulated CSL is stabilized due to the competition between field-induced commensuration and the chirality-protected helical ground state. A schematic diagram displayed in Fig. 1(b) presents a simple model of the evolution of the spin uniformity or coherence ($\xi$) as a function of magnetic field, $H$. [25] At $H = 0$, the spin coherence of the CHM phase is theoretically infinite due to the uniformity of the structure over the entire crystal. As $H$ is increased, $\xi$ abruptly disappears, marking the crossover into a distorted helicoid state. In this helicity-rich regime of the CSL, the growth of $\xi$ is minimal in a field range, $0 < H < H_{C,1}$. At a crossover field, $H_{C,1}$, $\xi$ increases more rapidly as the periodic ferromagnetic domains of the CSL grow and eventually diverge at the incommensurate to commensurate (IC-C) phase transition into the forced ferromagnetic (FFM) phase at $H_{C,2}$. The long-range coherence exhibited in the incommensurate CSL leads to special consequences such as collective dynamics, the character of which depends on the excitation frequency. [6,17,25] For example, in the microwave range, phonon-like modes or sliding dynamics of the CSL may occur at on- or off-resonance frequencies, respectively. [17]

On the other hand, the study of magnetization dynamics at frequencies much lower than the microwave range, ~ 0.1 – 10,000 Hz, can reveal the time-dependent response on length-scales that range from the level of the magnetic superlattice to individual moments of ferromagnetic domains. The time-dependent magnetization due to an alternating field, $H_{ac} = h \sin(\omega t)$, can be expanded as, [26]

$$M(t) = M_{1\omega} \sin(\omega t + \theta_{1\omega}) + M_{2\omega} \sin(2\omega t + \theta_{2\omega}) + M_{3\omega} \sin(3\omega t + \theta_{3\omega}) + ..., \qquad (1)$$

where $\omega = 2\pi f$, $M_{n\omega}$ is the $n^{th}$ harmonic component (for integer $n$ = 1, 2, 3, …), and $\theta_{n\omega}$ is the delay in phase of each component against $H_{ac}$. $M_{1\omega}$ represents the linear response to $H_{ac}$ while the non-zero contributions of the higher harmonics represent the degree of nonlinearity in the response of the magnetic system, i.e. the distortion of the periodic curve from sinusoidal behavior. Tsuruta, et al. [25] investigated the domain dynamics of the CSL across the phase boundaries separating the paramagnetic (PM) state for frequencies $f$ = 0.1 – 500 Hz and fixed dc fields, $H_{dc}$. At small $H_{dc}$, the dynamic response lacked the signature magnetic loss and third-order harmonic response ($M_{3\omega}$) attributed to the formation of ferromagnetic domains. This identified the PM-linear CSL transition. Conversely, as $H_{dc}$ was increased to a value close to $H_{C,2}$, the magnetic response exhibited large magnetic loss and $M_{3\omega}$ as a function of temperature on crossing into the highly coherent ferromagnetic domain-rich CSL from the FFM phase. This field regime of the CSL is referred to as the highly nonlinear (HNL) CSL.

Although the results in [25] allowed the delineation of the phase boundaries between the high temperature disordered state and the chiral magnetic phase below $T_0$, it is not clear how the dynamics change as a function of magnetic field as the CSL continuously evolves from linear regime and across the highly nonlinear regime. In general, the behavior of the magnetic response of long-wavelength structures includes contributions from spins on atomic length-scales and from the macroscopic spin structure. [27–29] These contributions have vastly different time-scales in which the latter may not be completely tracked by the ac measurement. Thus, the observation of the frequency dependence of the ac magnetic response may be used to identify the onset of the collective response of a spin structure that is coherent over large length-scales.

The schematic $H$-$T$ phase diagram for $Cr_{1/3}NbS_2$, presented in Fig. 1(c), summarizes experimental [16,25,30] and theoretical [31,32] results and depicts the phase boundaries near the critical temperature. An isothermal line at temperatures below the tricritical point, $T_{TCP}$, tracks the continuous transformation that coincides with the coherence model in Fig. 1(b). The IC-C phase boundary (black line), below which the chiral magnetic state exists, terminates at $T_0$ – the zero-field critical temperature that marks the onset of the CHM phase. According to recent experimental and theoretical studies, the spatially modulated phase is stable in a region above the Curie temperature, $T_C$. [30–32] In this region, a tricritical point along the chiral phase line separates the second-order HNL CSL-FFM transition from the first-order linear CSL-PM phase transition.

In [30], we reported phase boundaries that were carefully determined using the temperature and magnetic field dependence of the magnetic entropy change ($\Delta S_M$ ($T$,$H$)) and dc magnetization. Our results demonstrated that the chiral magnetic phase is stable above the Curie temperature within a precursor region ($T_C$ – $T^*$) analogous to the fluctuation-disordered regime observed in the cubic CHMs. [33,34] Although the first- and second-order behaviors were demonstrated experimentally, the location of the tricritical point was not experimentally resolved. The ac magnetic response reported in [25] identified a possible tricritical point separating the linear and nonlinear regimes, however, the results do not show refined detail in this region. As relaxation times observed at the phase boundaries in chiral helimagnets have been shown to decrease dramatically (over several orders of magnitude) at temperatures close to the phase transition [35], a wider frequency range may allow the refinement of distinctly different magnetic regimes to locate the tricritical point as well as detect the CSL behavior within the precursor region, $T_C - T^*$.

The following study presents an analysis of the ac magnetic response as a function of applied dc magnetic field, $M_{n\omega}(H)$, which tracks the dynamic response across the spatially modulated chiral phase into the FFM state. The measurement is taken across a logarithmic frequency range, $f = 11 - 10,000$ Hz, on a single crystal sample with $T_C = 130.75$ K. [30] We first present the field dependence of the linear ac magnetic response in Sec. III A. Sec. III B. follows with an analysis of the higher harmonic components of the ac magnetic response to clarify how the nonlinear response evolves as the chiral magnetic phase undergoes multiple field-induced crossovers. The study of the relaxation behavior of the HNL CSL is presented in Sec. III C. The temperature dependence, $M_{n\omega}(T)$, is presented in Sec. IV and clarifies the nature of correlations above the magnetic ordering temperature. Finally, in Sec. V, the phase diagram is constructed and the tricritical point is identified where the separation between the linear and highly nonlinear CSL regimes becomes apparent. A brief summary is given in Sec. VI.

## II. EXPERIMENTAL METHODS

The $Cr_{1/3}NbS_2$ single crystal was grown by a chemical vapor transport method that has been described elsewhere. [36] Measurements of the ac magnetic response, $M_{n\omega}$, were carried out using an AC Magnetometry System (ACMS) option for a commercial Physical Property Measurement System (PPMS, Quantum Design). Both the temperature and field dependence of $M_{n\omega}$, $M'_{n\omega}$, and $M''_{n\omega}$ were measured up to the fifth harmonic for the driving field $H_{ac} = h \sin(2\pi f$

$t$) with amplitude, $h$ = 5 Oe. The temperature-dependent ac magnetic response, $M_{n\omega}(T)$, was measured with a series of fixed dc fields ranging from $H_{dc}$ = 0 – 1200 Oe, parallel to $H_{ac}$, with a linear frequency range, $f$ = 10 – 10,000 Hz. The ac magnetic response as a function of applied dc magnetic field, $M_{n\omega}(H)$, was measured for a series of temperatures in the range $T$ = 129 – 133 K with a logarithmic frequency scale, $f$ = 11 – 10,000 Hz. Temperature- and field-dependent measurements were taken using a zero field-cooled (ZFC) protocol with $H_{ac/dc} \perp c$. A warming protocol was used between each measurement in which the sample was heated to 200 K, well above $T_C$, to remove history effects.

The complex ac magnetic response can be separated into real and imaginary parts,

$$M_{n\omega} = M'_{n\omega} + iM''_{n\omega}, \qquad (2)$$

where the imaginary component, $M''_{n\omega}$, is 90° out-of-phase with $M'_{n\omega}$ and reflects the presence of hysteresis or loss in $M(H_{ac})$.

## III. AC MAGNETIC RESPONSE: FIELD DEPENDENCE

### A. Linear response

Figs. 2(a) and (b) show the real and imaginary components of the linear magnetic response as a function of applied dc magnetic field, $M'_{1\omega}(H)$ and $M''_{1\omega}(H)$, measured with $f$ = 111 Hz for selected temperatures ranging between $T$ = 129 – 133 K. While $M'_{1\omega}$ is non-zero for all fields measured, $M''_{1\omega}$ appears only at 0 Oe and in a field range between $H$ = 250 – 570 Oe, i.e. where $M'_{1\omega}$ shows a giant response. In the context of the simple spin coherence model described in Sec. I, the following describes the behavior of the ac magnetic response as the magnetic system continuously transforms from a spatially-modulated chiral phase to the homogenous FFM state: At $H = 0$ Oe, $M'_{1\omega}$ displays a maximum, which is accompanied by a large $M''_{1\omega}$ due the essentially infinite spin coherence of the CHM structure, which is spatially uniform over the entire crystal. As magnetic field is increased in steps of 10 Oe, $M'_{1\omega}$ drops to a relatively constant value. $M''_{1\omega}$ abruptly goes to zero and marks the crossover into the linear CSL. In this regime, long-range spin coherence disappears, extending over a broad field range from $\sim H = 30 - 250$ Oe for $T = 129$ K. At a critical field, the spin system crosses over into a HNL CSL in which both $M'_{1\omega}$ and $M''_{1\omega}$ show an exponential-like increase that evolves into a broad peak with sharp anomalies. Above a critical magnetic field, $M'_{1\omega}$ drops to a minimum and loss disappears marking the IC-C phase transition into the FFM phase.

Figs. 2(c) and (d) show the $H$-$T$ surface plots of $M'_{1\omega}(H,T)$ and $M''_{1\omega}(H,T)$ for $f = 1111$ Hz. The double anomalies of the HNL CSL are apparent as dark red ridges. The large linear response and accompanying magnetic loss of the HNL CSL extend past the Curie temperature, $T_C = 130.75$ K measured for this system, [30] into a region marked by strong chiral correlations, in agreement with results in [30–32]. For temperatures above ~131.5 K, the magnetic loss vanishes. Here, $M'_{1\omega}$ remains as a single broad peak (see also Figs. 2(a) and (b)) and extends to $T_0 \sim 132.25$ K, which marks the disappearance of the chiral magnetic phase. This feature without magnetic loss emerges on the boundary between the linear CSL and PM states. [25] The $H$-$T$ surface plots mirror the phase diagram shown in Fig. 1(b). Namely, the HNL CSL-FFM transition and linear CSL-PM transitions meet in the region $T_C - T_0$.

While the behavior of the domain dynamics elucidated the HNL CSL-FFM and linear CSL-PM phase boundaries as a function of temperature in [25], the nature of the field-dependent crossover from the helicity-dominated linear regime of the chiral phase into the highly coherent ferromagnetic domain-rich nonlinear regime of the CSL is still unclear. To understand the evolution of the spatially-modulated chiral phase, the study begins with a comparison of the static (differential) and dynamic susceptibilities. Fig. 3 compares the differential susceptibility, d$M$/d$H$, shown by the black curve, and the ac susceptibility, $M_{1\omega}'/h$, shown by the green curve, for $T$ = 129 K at the lowest frequency (longest time-scale) measured in this study, $f$ = 11 Hz. d$M$/d$H$, derived from dc magnetization versus magnetic field measurements, shows a single peak at a characteristic field, $H_{dM/dH}^{peak}$. As the susceptibility reaches its maximum, in a field regime where the coherent FM domains dominate the magnetic structure, the dynamic susceptibility deviates from the behavior of d$M$/d$H$, splitting into two anomalies of lesser magnitude. The location of $H_{dM/dH}^{peak}$ lies exactly at the position of the minimum between the two peaks in $M_{1\omega}'/h$. A closer inspection indicates that the deviation between the dc and ac susceptibilities becomes prominent near $H$ = 250 Oe and coincides with the onset of non-zero magnetic loss, shown by the orange curve. Clearly, the dynamic susceptibility at 11 Hz, which has a time window of ~ 90 ms, does not track the susceptibility measured in the static limit. The dynamic response is instead suppressed in magnitude and is accompanied by an increase in magnetic loss.

The discrepancy in the magnetic field dependence of the susceptibility has been observed in many systems with long-wavelength magnetic structures, e.g. CHM and SkL phases. [4,29,37] The difference in magnitude has been linked to (1) a slow field-driven process, such as the reorientation of large helical domains, that may only be fully observed in the zero frequency limit, and (2) phase coexistence and strong dissipation accompanying a first-order transition. [29] As the field-driven crossover in $Cr_{1/3}NbS_2$ is a continuous process, [16] the dynamic behavior shown here is characteristic of long-wavelength magnetic structures that relax on macroscopic time-scales. [27] Indeed, the suppression of the susceptibility is enhanced with higher frequency and will be explored further in Sec. III. C. In the following section, we will examine the variation of the nonlinear response in the field regime marked by slowing dynamics to identify the onset of collective dynamics of a coherent macroscopic spin state.

**B. Nonlinear response**

We begin with a brief review of nonlinear ac magnetic response in magnetic systems. The relative magnitudes of higher harmonic components represent the distortion of the time-dependent magnetization from typical sinusoidal behavior in response to an ac driving field, $H_{ac} = h\sin(\omega t)$. Mito and coworkers have rigorously studied the nonlinear response of chiral magnetic materials [38–41], mainly as a tool to study magnetic domain dynamics: The dynamical magnetization, $M(H_{ac})$, with large $M_{3\omega}$ and magnetic loss is modelled using the nonlinear spring (Duffing) equation to describe the displacement of domain walls from equilibrium. [40] In light of the description of the collective dynamics of the chiral soliton lattice in relation to a modified Duffing oscillator model [20], Tsuruta, et al. applied this technique to $Cr_{1/3}NbS_2$ to elucidate the dynamics across different regimes of the CSL with increasing temperature. [25] Beyond domain dynamics, early studies of higher-order susceptibilities yielded descriptions of the phase transitions in canonical systems, i.e. FM, antiferromagnet, and spin glass [42,43]: The third harmonic identifies the nature of the magnetic ordering at the phase transition. It is linked to the breaking of spatial-reversal symmetry and yields information about the spin environment. [38,39] Additionally, the even harmonics are dependent on the presence of a symmetry-breaking internal field and are commonly utilized as an unambiguous detection of spontaneous magnetization. [44–46] Thus, the analysis of the nonlinear ac magnetic response can be a powerful tool to understand the fundamental phenomena across various magnetic regimes.

The first five harmonics of the ac magnetic response as a function of magnetic field, $M_{n\omega}(H)$, at $f = 111$ Hz are compared in Figs. 4(a)-(f), where the black curves represent the measurement at $T = 129$ K. The first column, Figs. 4(a)-(c), compares the large response of the higher-order odd harmonic components, $M_{3\omega}(H)$ and $M_{5\omega}(H)$, to the magnetic loss $M''_{1\omega}(H)$. In Figs. 4(e) and (f), the large responses of the even harmonic components, $M_{2\omega}$ and $M_{4\omega}$, are displayed and show a complex modulation across the highly nonlinear regime. As a reference between even and odd harmonics $M'_{1\omega}$ is displayed in Fig. 4(d).

$M_{2\omega} - M_{5\omega}$ capture the disappearance of spin uniformity which is predicted to occur as the CHM structure crosses over into the CSL. At $H = 0$ Oe, the magnetic response of the CHM structure displays a contribution from all higher harmonics, $M_{2\omega} - M_{5\omega}$, and abruptly drops at small $H$. The minimum, which extends for a relatively wide field range, rapidly rises at a characteristic field, $H_{C,1}$.

As shown by the black dashed lines through Figs. 4(a)-(c), $M_{3\omega}(H)$ and $M_{5\omega}(H)$ display virtually the same field dependence as $M''_{1\omega}(H)$: The onset of $M''_{1\omega}$ is accompanied by a rise in $M_{3\omega}$, and appearance of $M_{5\omega}$, and is followed by an exponential-like increase to sharp double anomalies. Thus, the onset of slowing dynamics, indicated by $M''_{1\omega}$ (Fig. 3), corresponds to a prominent nonlinear response which, in the picture of magnetic domain dynamics, is associated with the growth of the FM domain component of the CSL. The third harmonic response as a function of field, $M_{3\omega}(H)$, for $f = 25 - 10,000$ Hz is shown in Fig. 5 for $T = 129$ K. As frequency is varied, the regime of large $M_{3\omega}$ remains rigid as a function of magnetic field and shows no shift of the sharp double anomaly (Fig. 5). $M_{3\omega}$ displays the characteristic dip in magnitude that was observed in $M'_{1\omega}(H)$ and $M''_{1\omega}(H)$, which corresponds to peak in the static susceptibility, $H_{dM/dH}^{peak}$.

While the highly nonlinear regime of the CSL was previously characterized [25] with respect to the large $M_{3\omega}$, the large magnitude of the higher harmonics from $M_{2\omega}$ to $M_{5\omega}$ reinforces that the time-dependent magnetization is highly distorted in response to the sinusoidal driving field. The relative magnitudes of the higher harmonics to $M_{1\omega}$ further demonstrate the high nonlinearity in the system. Specifically, the ratio of $M_{3\omega}/M_{1\omega}$, called the Klirr factor, is used as a measure of the nonlinearity and was reported previously as approximately 10% in this system. [25] The measurements reported herein demonstrate a large Klirr factor of up to 15%. It is also worth noting that $M_{5\omega}$ is in the range of 16% of $M_{3\omega}$. Depending on the measurement frequency, the peak magnitude of $M_{2\omega}$ is comparable to $M_{3\omega}$, as seen in Fig. 4(e) where $M_{2\omega}(H)$ measured at $T = 130.75$ K reaches a magnitude of ~ 1.3 (emu/mol). Consequently, the even order terms of (1) contribute a considerable distortion to the time-dependent magnetization and, similarly to $M_{3\omega}$ and $M_{5\omega}$, reflect the collective response of the coherent FM domains of the HNL CSL.

The magnetic field dependence of the second harmonic magnetic response, $M_{2\omega}(H)$, is shown in Fig. 6(a) for $T = 130.5$ K. As observed in the odd order ac magnetic responses, $M_{2\omega}$ displays a complex field dependence across the HNL CSL, which is further demonstrated by the real and imaginary components ($M'_{2\omega}$ and $M''_{2\omega}$) shown in the inset. Both the in-phase and out-of-phase components contribute almost equally to the total $M_{2\omega}$, which signifies significant energy loss associated with the changes in internal field. Due to the presence of spontaneous magnetization in a system, there exists an asymmetry in the magnetization with respect to the direction of the applied magnetic field in an ac measurement. [44], [45] Furthermore, sharp anomalies in $M_{2\omega}$ should accompany sudden changes in internal field. [44], [45] The $M_{2\omega}$ response is largest upon entering and exiting the highly nonlinear regime as field increases, where on entry a sharp positive to negative sign change occurs and on exit, small positive values sharply switch to negative moving into the FFM phase. The sign changes in $M'_{2\omega}$ and $M''_{2\omega}$ across the highly nonlinear regime can be observed for all temperatures measured up to $T \sim 131.5$ K, as can be seen in the $M'_{2\omega}(H,T)$ surface plot in Fig. 6(b). $M_{2\omega}(H)$ displays sudden changes associated with the onset of the rich ferromagnetic domain component of the HNL CSL. This may be associated with a precipitous increase in ferromagnetic domain size which, based on static magnetization measurements, thereafter rapidly increases with magnetic field. [47] To our knowledge the magnitude of $M_{2\omega}$ is usually not a significant contribution to the total measured ac magnetic response, and therefore the results have not been formally displayed in previous studies of the nonlinear response of chiral magnets. [25,40]

The field dependence of $M_{2\omega} - M_{5\omega}$ clearly portrays the evolution of the CSL from one regime to another: the linear CSL, which displays a predominantly in-phase linear response to an ac magnetic field, and the HNL CSL, which includes large nonlinear contributions and large magnetic loss to the total response. The attenuation in magnitude of $M_{1\omega} - M_{5\omega}$ at $H_{dM/dH}^{peak}$ is a signature of the excessive slowing of the dynamics across the highly nonlinear CSL as the FM domains continuously grow with increasing $H$. [25] To further understand the relaxation behavior across the highly nonlinear regime, we analyze the frequency dependence in the following section.

## C. Frequency dependence

The frequency dependence of $M'_{1\omega}(H)$ and $M''_{1\omega}(H)$ (Figs. 7(a) and (b)) further illustrates the effect of a decreasing time window as the magnitude of $M'_{1\omega}$ decreases with frequency. To investigate the effect of frequency in more detail, a quantitative analysis of the frequency dependence of the linear susceptibility, $\chi = M_{1\omega}/h$, is possible from the Cole-Cole modification [48] of the Debye model,

$$\chi(\omega) = \chi(\infty) + \frac{\chi(0) - \chi(\infty)}{1 + (i\omega\tau_0)^{1-\alpha}}, \qquad (3)$$

which introduces the parameter $\alpha$ to account for a distribution of relaxation times: $\alpha = 1$ corresponds to an infinitely broad distribution and $\alpha = 0$ accounts for a single relaxation process. The Cole-Cole model assumes a distribution of relaxation times which is symmetric about $\tau_0 = 1/(2\pi f_0)$, the characteristic or average relaxation time, on a logarithmic scale. Here, $\chi(0)$ and $\chi(\infty)$ are the isothermal and adiabatic susceptibilities, which correspond to spin-lattice and spin-spin interactions, respectively. $\chi(\omega)$ can be decomposed into in- and out-of-phase components,

$$\chi'(\omega) = \chi(\infty) + \frac{(\chi(0) - \chi(\infty))\left[1 + (\omega\tau_0)^{1-\alpha}\sin(\pi\alpha/2)\right]}{1 + 2(\omega\tau_0)^{1-\alpha}\sin(\pi\alpha/2) + (\omega\tau_0)^{2(1-\alpha)}} \qquad (4)$$

$$\chi''(\omega) = \frac{(\chi(0) - \chi(\infty))(\omega\tau_0)^{1-\alpha}\cos(\pi\alpha/2)}{1 + 2(\omega\tau_0)^{1-\alpha}\sin(\pi\alpha/2) + (\omega\tau_0)^{2(1-\alpha)}}. \qquad (5)$$

Thus from fits of (4) and (5) to the frequency dependence of the real and imaginary components of the linear susceptibility, the above parameters can be extracted to analyze the change in the dynamics as a magnetic phase evolves with temperature and magnetic field.

Figs. 7(c) and (d) plot the real and imaginary components of $\chi_{1\omega}$, respectively, as a function of frequency at $H = 490$ Oe, which corresponds to $H_{dM/dH}^{peak}$ at $T = 129$ K. $\chi''_{1\omega}$ displays an asymmetry about its peak which corresponds to the characteristic frequency $f_0$ or, equivalently, the average relaxation time, $\tau_0$ (Fig. 7(d)). Additionally, $\chi''_{1\omega}$ approaches an apparent non-zero value in the isothermal limit, $\chi(\omega \to 0)$. The solid lines in Fig. 7(c) and (d) demonstrate how the model deviates from the measured $\chi'_{1\omega}$ and $\chi''_{1\omega}$, where separate fits were performed on either side of the peak at $\tau_0$. To account for the low frequency behavior of $\chi''_{1\omega}$, following a similar procedure reported in [49], an additional frequency-independent term, $\chi_0''$, was added to the right hand side of (5).

Fig. 8(a) plots the evolution of $\tau_0$ with temperature for selected fields within the HNL CSL regime. For each magnetic field, an acceleration of the dynamics is observed as the characteristic time drops with an increase in temperature, varying from $\sim 10^{-3} - 10^{-5}$ s approaching $T = 131.5$ K. However, as $H$ is increased and the FM domain component of the HNL CSL grows, a general slowing trend is observed between curves, most noticeably as magnetic field increases past $H_{dM/dH}^{peak}$.

As demonstrated by the fits to $\chi'_{1\omega}$ in Fig. 7(c), the asymmetry is less pronounced with respect to the inflection point at $\tau_0$. However, at the lowest frequencies, $\chi'_{1\omega}$ deviates significantly from the expected sigmoidal dispersion. In this regime, the susceptibility displays an almost linear increase in magnitude as frequency varies from 39 – 11 Hz. Thus, as the measurement window approaches macroscopic time-scales, an additional dynamic process is captured by both the real and imaginary components of the susceptibility. Fig. 8(b) shows the trend in the constant term $\chi''_0$, which gradually drops with temperature to a value close to zero at $T = 131.5$ K. The drop in $\chi''_0$ may indicate that the gradual loss of a competing dynamic process which disappears above the tricritical point.

Anomalous relaxation phenomena with respect to the Cole-Cole model has been observed in other magnetic systems with spatially-modulated structures such as the cubic chiral helimagnets and a cycloidal magnet, $GaV_4S_8$. [37,49,50] In each of these systems the behavior of the deviation varies: The frequency dependence of $\chi''_{1\omega}$ in $Fe_{1-x}Co_xSi$ displays a similar profile to the data presented in this study and is attributed to the coexistence of multiple phases due to chemical doping. [50] In [49], it was reported for $Cu_2OSeO_3$ that the behavior of $\chi''_{1\omega}$ implied a symmetric distribution of relaxation times, yet a frequency-independent component, $\chi''_0$, was added to the model for the full frequency range. In $GaV_4S_8$, separate fits of the Cole-Cole model to $\chi'_{1\omega}$ and $\chi''_{1\omega}$ failed to produce the same parameter set. [37] Unlike in $Cr_{1/3}NbS_2$, where the anomalous behavior is representative of the dynamics of a pure magnetic phase, the behavior in these systems is observed on phase boundaries between modulated phases. Nevertheless, the relaxation phenomena in each of these systems are complicated by the slow dynamics of magnetic structures on large length-scales.

The acceleration of the dynamics with increasing temperature is also clear in the frequency dependence of the real and imaginary components of the third harmonic response, $M'_{3\omega}$ and $M''_{3\omega}$, Figs. 9(a)-(d). As previously discussed in Sec. III. B., the field range of the large $M_{3\omega}$ signal is relatively rigid as frequency is varied. However, the character of $M'_{3\omega}$ and $M''_{3\omega}$ varies significantly with frequency across the HNL CSL. In Fig. 9(a), $M'_{3\omega}$ starts off as a large negative value that gradually reduces in magnitude with increasing frequency. For $f \sim 111 - 155$ Hz, $M'_{3\omega}$ exhibits a negative to positive crossover. Above 155 Hz, $M'_{3\omega}$ increases to a positive peak. The gradual evolution of the sign change in $M'_{3\omega}$ is typically attributed to the ac measurement probing the response of spin interactions and other degrees of freedom at different time, and hence, spatial scales. [39,43,46] The variation in the characteristic frequency dependence described above is summarized in Figs. 9(b)-(d), where the location of the inflection point of the negative (blue) to positive (red) crossover in $M'_{3\omega}(H,T)$ shifts to higher temperatures with increasing frequency.

The decrease in time-scale at progressively higher temperatures has been observed in magnetic systems with long-wavelength structures and suggests the thermal activation of relaxation processes. [35,37] However, the behavior may not always be explained in terms of a simple Arrhenius model, $f_0 = A\exp(-E_a/k_BT)$, which can lead to unphysically large energy barriers. [49] Fig. 10 plots the extracted parameters as $\ln f_0$ vs $1/T$. The inset illustrates the acceleration of the dynamics as the peak in $\chi''_{1\omega}$ vs $f$ shifts to higher values of $f_0$ as temperature increases toward the phase transition at $T \sim 131.5$ K. The behavior of spin relaxation clearly falls outside of a simple thermal activation scheme in which a linear trend in $\ln f_0$ vs $1/T$ would be expected. However, the data show a clear trend in the dynamic response, which speeds up on approaching the tricritical point. The frequency regime used in this work allows access to shorter time-scales than in previous studies, which may aid in refining the destruction of the highly nonlinear regime to identify the tricritical point.

## IV. AC MAGNETIC RESPONSE: TEMPERATURE DEPENDENCE

To gain a comprehensive portrait of the phase diagram, we explore the behavior of the even and odd higher harmonic responses with temperature across several magnetic field regimes. As demonstrated in Sec. III C., the dynamic phenomena close to the phase transition gradually approach shorter time-scales. Therefore, a particular emphasis is placed on measurements at the higher end of the frequency spectrum to clarify the phase evolution within the precursor region, $T_C - T^*$.

Fig. 11 shows the temperature dependence of $M_{1\omega} - M_{5\omega}$ of the CHM phase at 0 Oe and $f$ = 10,000 Hz. The familiar sharp kink in $M_{1\omega}$ corresponds to the phase transition into the CHM phase from the PM state [10,30,51,52] at $T_0 \sim 132.25$ K. The inflection point, which is noted in other chiral helimagnets to mark a fluctuation-disordered precursor region [33,34], occurs at $T^*$ = 133 K and is in close agreement with our $M$ vs $T$ results reported in [30]. The inflection point at $T^*$ coincides with the onset of strong $M_{2\omega}$ and $M_{3\omega}$, which supports the predictions of chiral correlations in the precursor region and suggests an increasing coherence of fluctuations as the phase transition is approached. In fact, a large zero-field $M_{3\omega}(T)$ has been demonstrated to detect the effects of crystalline chirality on magnetic correlations above the transition temperature. [38]

The temperature dependence of $M_{n\omega}$ at fixed dc fields is presented in Figs. 12(a)-(d). Figs. 12(a) and (b) show the temperature dependence of the real and imaginary components of the linear magnetic response, $M'_{1\omega}(T)$ and $M''_{1\omega}(T)$, measured in this study at $f = 100$ Hz and with dc field, $H_{dc} = 50 - 1200$ Oe. The two field-dependent anomalies in $M'_{1\omega}(T)$ at high and low temperature, respectively, are consistent with results reported by Tsuruta, et al. in [25]. Namely, for $H_{dc} > 400$ Oe, a shallow peak emerges in $M'_{1\omega}$ at high temperature, $T = T_m$ ($> T_C$), that is not accompanied by loss. A similar feature has been observed above $T_C$ in ac magnetic measurements of the cubic chiral helimagnets and is associated with the transition from the paramagnetic into the field-polarized (FP) state. [29,33] It has already been demonstrated that strong ferromagnetic correlations exist well above $T_C$ in $Cr_{1/3}NbS_2$ [30] and in other CHMs. [34,53] Thus, the transition at $T_m$ indicates the temperature regime in which the FM correlations become strongly interacting. In this case, under an applied magnetic field, $M_{2\omega}(T)$ should be present. The second harmonic magnetic response displayed in the inset supports this picture, where the peak in $M_{2\omega}(T)$ coincides with the peak at $T_m$ in $M'_{1\omega}(T)$ measured at 1200 Oe. $T_m$ shifts to higher temperature as magnetic field is increased and Zeeman energy stabilizes FM correlations at higher temperature.

For $H_{dc} \geq 400$ Oe, the low temperature peak in $M'_{1\omega}(T)$ is accompanied by a significant response in $M''_{1\omega}(T)$ which, as established in [25], reflects the energy loss of the ferromagnetic domains of the HNL CSL against the time-dependent field. Similar to the behavior of $M'_{1\omega}(H)$ reported in Sec. III A., the center of the broad peak exhibits a small dip in magnitude and coincides with the center of the highly nonlinear regime of the CSL at $H_{dM/dH}^{peak}$. The abrupt disappearance of magnetic loss as temperature is increased, occurs at phase the boundary between the HNL CSL and FFM phase. The low temperature peak in $M'_{1\omega}(T)$ becomes suppressed at lower temperatures as the dynamic processes of the HNL CSL slow significantly moving away from the tricritical temperature. Specifically, at $H_{dc} = 800$ Oe, the $M'_{1\omega}(T)$ maximum is completely suppressed, yet the domain dynamics are still apparent from the non-zero $M''_{1\omega}(T)$ down to $T \sim 115$ K in Fig. 12(b). The same behavior is observed in [25], where it was noted that the anomaly in $M'_{1\omega}(T)$ measured at $f = 1$ Hz is temporarily enhanced and then suppressed at lower temperature. In the present study, the measurement of $M'_{1\omega}(T)$ for $f = 100$ Hz accesses time-scales that are faster by 2 orders of magnitude. Hence, the regime of maximum response occurs at temperatures closer to the phase transition.

Figs. 12(c) and (d) demonstrate the temperature dependence of $M_{2\omega}$ and $M_{3\omega}$ for low magnetic fields, $H = 50, 200, 400$ Oe, measured at $f = 10,000$ Hz. A double peak in $M_{2\omega}$ gradually develops with successively higher fields up to 400 Oe, where the anomaly is accompanied by a large $M_{3\omega}$ of the HNL CSL. The $M_{2\omega}$ response at 50 Oe and 200 Oe which lack a giant $M_{3\omega}$ and a magnetic loss signature in Fig. 12(b) represents the PM-linear CSL transition. Hence, the transition from PM-linear CSL displays a nontrivial change in internal field with a similar character to the transition into the HNL CSL. The non-zero $M_{2\omega}$ signal indicates the presence of small field-polarized regions in an otherwise helicity-rich structure. It is likely due to the gradual formation of short-range ferromagnetic regions with increasing magnetic field, which agrees with the theoretical picture in which the spatial period of the CSL continuously grows as $H$ increases.

## V. PHASE DIAGRAM AND TRICRITICAL POINT

In this section, a comprehensive phase diagram is constructed to summarize the features of the field and temperature dependence of the crossovers and phase transitions in $Cr_{1/3}NbS_2$. First, the details of the determination of the critical values are presented in Fig. 13. Fig. 13(a) directly compares the field dependence of the first derivatives of the linear and higher order magnetic responses, $dM'_{1\omega}/dH$, $dM_{3\omega}/dH$, $dM_{2\omega}/dH$ for the measurement at $T = 129$ K. The magnetic responses all display sharp changes in slope across the highly nonlinear regime of the CSL. The collapse in the spin coherence of the CHM state with an increase of magnetic field from $H = 0$ Oe is observed in all measurements, after which the change in slope is minimal over a broad field range. Both $dM'_{1\omega}/dH$ and $dM_{3\omega}/dH$ display a gradual increase near $H = 300$ Oe. This coincides with the apparent onset of the frequency dependence in $M_{1\omega}$ and $M_{3\omega}$ as well as the deviation between the static and dynamic susceptibilities and appearance of magnetic loss (Fig. 3). Above 400 Oe, rapid jumps in slope occur in $M_{1\omega}$, $M_{3\omega}$ and $M_{2\omega}$, which we define as the crossover field, $H_{C,1}$. In the field regime above $H_{C,1}$, the CSL is characterized by the onset of an anomalous $M_{3\omega}$ response which coincides with extremely slow dynamics associated with the collective response of spatially coherent FM domains. This behavior falls off at $H_{C,2}$, the critical field for the IC-C transition into the FFM phase.

The determination of the critical temperature, $T_m$, which separates the paramagnetic and field-polarized state is shown in Figure 13(b). As demonstrated in Fig. 12(a), $T_m$ shifts to higher temperature with increasing magnetic field. $M'_{1\omega}(T)$ for $f = 10,000$ Hz measured in the vicinity of the tricritical point are displayed for fixed dc fields ranging from $H_{dc} = 400 - 550$ Oe. Unique $T_m$ values were extracted from a peak fit of each curve, as demonstrated in the inset. The disappearance of $T_m$ indicates the location of the tricritical temperature, below which the continuous transformation of the chiral helix into the ferromagnetic state is achieved via a crossover into a highly nonlinear CSL.

Figs. 14(a)-(f) display the surface plots of $M'_{1\omega} - M'_{5\omega}(H,T)$ and $M''_{1\omega}(H,T)$ measured at $f = 10,000$ Hz, for which the regime of maximum ac magnetic response runs from $T_C$ to $T_{TCP}$. Critical field and temperature values determined from anomalies in $M_{n\omega}$ are superimposed. In Fig. 14(a), $M'_{1\omega}$ demonstrates the extent of the chiral magnetic phase down to $T_0$, the zero-field critical temperature of the PM-CHM transition. The chiral phase extends past the ferromagnetic Curie temperature calculated for this system [30] into a precursor region, $T_C - T^*$. In all plots of $M_{n\omega}$, a strong response is observed near $H = 0$ Oe, reinforcing the conclusion of Tsuruta, et al. [25] that the CHM phase exists as a singularity in the absence of applied magnetic field. The large response is present up to $T^* = 133$ K as defined in Sec. IV., which may signify chiral correlations in a precursor regime analogous to the fluctuation-disordered region observed in the cubic chiral helimagnets. [33,34]

The critical fields, $H_{C,2}$, at which $M'_{1\omega}$ and $M''_{1\omega}$ fall off are marked by open black symbols below $T_{TCP}$ and in red for $T > T_{TCP}$. A simple power law fit to the IC-C phase line is given by $H_{C,2} \propto (T - T_0)^{0.258 \pm 0.031}$ with $T_0 = 132.3 \pm 0.05$ K (solid black line). $H_{C,1}$ and $H_{dM/dH}^{peak}$ are tracked by dashed lines extrapolated to $H = 0$ Oe using a similar power law to $H_{C,2}$. The PM-FP line is given by the location of $T_m$, determined from peak fits of $M'_{1\omega}(T, H_{dc})$ (green squares) as demonstrated in Fig. 13(b). $T_m$ values obtained from $M'_{1\omega}(T, H)$ reformulated from field-dependent data (hollow pink squares) are in good agreement with temperature-dependent data, which demonstrates negligible hysteresis across the field-polarized transition. The intersection of the PM-FP phase line with $H_{C,2}$ at $T_{TCP}$ = 131.35 K defines the tricritical point that separates the HNL CSL-FFM and the linear CSL-PM transitions.

Fig. 14(b) labels the magnetic phases and mirrors the schematic phase diagram presented in Fig. 1(c) in Sec. I. The temperature and field dependence of each harmonic, Figs. 14(b)-(f), illuminates the destruction of the HNL CSL above the tricritical temperature, as the dynamic signatures of the periodic array of ferromagnetic domains disappear. Small $M_{2\omega}$ and $M_{3\omega}$ values exist above $T_{TCP}$ at the PM-linear CSL transition. As already demonstrated in the temperature dependence section, $M_{2\omega}$ points to the change in internal field due to the gradual formation of short-range FM regions as a precursor to the crossover into the HNL CSL regime. In general, the presence of non-zero $M_{3\omega}$ accompanying $M_{1\omega}$ is expected at a phase transition. [26] However, it lacks the large magnetic response due to the formation of large FM domains which are spatially coherent over large length-scales. The phase diagram calculated from dynamical measurements distinguishes the linear and HNL CSL regimes and reinforces the existence of the tricritical point that has been theoretically predicted. [31,32]

While the phase diagram based on the field dependence of $M_{n\omega}$ presented here focuses on a temperature regime close to the phase transition, the measurement window detects the signature magnetic loss of the FM domains for temperatures down to at least $T$ = 110 K, as demonstrated in Sec. IV. The dynamic behavior studied within the frequency range $f$ = 11 – 10,000 Hz reinforces the results of Tsuruta, et al. [25], measured at frequencies as low as 0.1 Hz. In the present study, $T_m$ shifts from $T$ = 131.35 – 135 K over fields $H$ = 0.3 – 0.8 $H_{C,0}$, where $H_{C,0}$ refers to the critical field extrapolated to absolute zero. In [25] the phase line also shifts with applied field, over a range of ~ 5 K as dc magnetic field varies from an estimated $H \sim 0.4 - 0.9$ $H_{C,0}$. The frequency range employed herein refines details of the phase diagram at high temperatures, where the dynamics are significantly accelerated.

The broad field range of the linear CSL regime (ranging between $0 < H < H_{C,1} = 410$ Oe at $T = 129$ K) is testament to the competition between the symmetry-protected chirality of the magnetic state and the external magnetic field which forces commensuration. This implies that the Zeeman energy must reach a critical value before the system crosses over into a ferromagnetic-domain dominated state, after which the growth of the commensurate regions presumably becomes more rapid with increasing magnetic field. The consequences of this competition are also seen in previous studies of the dc magnetization, which displays a linear growth at low field followed by a rapid nonlinear increase before reaching saturation. [25,30,36] Furthermore, as temperature is increased, the field regime becomes smaller as thermal disorder destabilizes the competition. Fig. 14(c) displays a line at low field that represents the deviation from linearity of the dc magnetization as a function of magnetic field curves, which was suggested as a possible crossover boundary between the linear CSL and HNL CSL in [25]. The measurements reported herein demonstrate that the system requires higher applied magnetic fields to exhibit the large nonlinearity in the magnetic response.

The initial growth of the magnetic loss, as seen in Fig. 3, is quite slow and precedes the onset of the enormous magnetic response by a relatively large field interval, $\Delta H \sim 250 - 410$ Oe. This behavior points to a gradual formation of commensurate regions, which agrees with the theoretical picture of a modulated CSL that evolves continuously from a simple chiral helix. At $H_{C,1}$, our measurement detects a rather sharp increase in the magnetic response and magnetic loss (Fig. 13(a)) as the system crosses over into the HNL CSL. However, it is important to emphasize that the theoretical description of the field-induced evolution from a simple spin helix into a modulated HNL CSL is a completely *continuous* process. The dynamic magnetic response presented in this report sensitively detects changes in the magnetic structure. [54] According to early neutron diffraction studies of the soliton lattice by Izyumov and Laptev [55], the scattering amplitudes representing the first-order (harmonically-modulated) component and the zero-order (ferromagnetic) component cross at a magnetic field below the critical field for the IC-C transition ($H < H_{C,2}$). [54] At the crossing point, the physics of the ferromagnetic domains may begin to dominate the dynamic response, [54] leading to anomalously large magnetic loss and nonlinear ac magnetic response.

## VI. SUMMARY

We investigated the magnetic field-driven crossovers of the incommensurate chiral magnetic structures in the monoaxial helimagnet $Cr_{1/3}NbS_2$ via the magnetic field and temperature dependence of the ac magnetic response. As magnetic field is increased perpendicularly to the chiral spin helix, the growth of the spatial period of the commensurate domains of the CSL is initially slow. However, at a crossover field, $H_{C,1}$, the FM domain component dominates the spin structure and marks the crossover into a highly nonlinear CSL. The anomalous ac magnetic response observed above $H_{C,1}$ sensitively detects this change in magnetic structure, which coincides with the onset of extremely slow dynamics.

The deviation in the static and dynamic susceptibilities of the HNL CSL is characteristic of a large magnetic structure that relaxes on macroscopic time-scales. An investigation of the frequency dependence of the susceptibility demonstrates that the dynamic response in the highly nonlinear regime of the CSL exhibits an asymmetric distribution of relaxation times. The dispersion and loss in the linear ac susceptibility indicate the presence of a competing dynamic process, which is gradually lost as the dynamics speed up with increasing temperature.

A thorough investigation of the $M_{2\omega}$ component of the nonlinear response has been presented for the first time. $M_{2\omega}$ probes the changes in internal field in both the HNL CSL and the linear CSL and exhibits signatures of spontaneous magnetization in both structures. This suggests a gradual increase in the spatial period of commensurate regions throughout the CSL, and agrees with the theoretical picture of simple chiral helix which continuously transforms into a homogenous FFM phase via a CSL.

Each harmonic, $M_{1\omega} - M_{5\omega}$, illuminates the destruction of the HNL CSL above the tricritical point. Based on the power law dependence of the IC-C phase line, $H_{C,2}$, and the determination of the paramagnetic to field-polarized transition, $T_m$, the tricritical point at $T_{TCP} =$ 131.35 K is experimentally resolved which separates the HNL CSL-FFM transition and the linear CSL-PM transition.

**ACKNOWLEDGMENTS**

Research at the University of South Florida was supported by the U.S. Department of Energy, Office of Basic Energy Sciences, Division of Materials Sciences and Engineering under Award No. DE-FG02-07ER46438. L.L. and D.G.M. acknowledge support from the National Science Foundation under grant DMR-1410428. Research at the Naval Research Laboratory was funded by the Office of Naval Research (ONR) through the Naval Research Laboratory Basic Research Program. We thank Professor Jun-ichiro Kishine of The Open University of Japan and Professor Alexander Ovchinnikov of Ural Federal University for useful discussions.

**Figure captions:**

FIG. 1. Evolution of the spatially-modulated chiral spin structure in $Cr_{1/3}NbS_2$ with applied magnetic field. (a) The CHM structure continuously evolves into a CSL. The period of the helical ground state is $L(0) = 48$ nm. The CSL period, $L(H)$, continuously grows with applied magnetic field, $H$, perpendicular to the $c$ axis. The forced ferromagnetic (IC-C) transition occurs at magnetic fields above $H_{FFM}$. (b) The change in spin coherence or uniformity of the magnetic structure as a function of magnetic field, $\xi(H)$, as described in the text. The CHM is coherent over the entire crystal at $H = 0$ Oe. In the CSL regime, the spin uniformity corresponds to ferromagnetic domain (commensurate) component of the magnetic structure. (c) Schematic $H$-$T$ phase diagram. The green line at $H = 0$ Oe represents the pure CHM state where $\xi(0) = \infty$. The PM-CHM transition occurs at $T_0$. The CSL is divided into two regimes by a crossover boundary at $H_{C,1}$ (blue line) which separates the linear and HNL CSL. The chiral phase boundary extends past the Curie temperature, $T_C$, and terminates at $T_0$. A precursor region of strong correlations is marked by $T^*$. The tricritical point, $T_{TCP}$, separates a second-order HNL CSL-FFM transition with a first-order linear CSL-PM transition.

FIG. 2. Magnetic field dependence of the real and imaginary parts of the linear ac magnetic response, $M_{1\omega}(H)$, measured with an ac magnetic field amplitude, $h = 5$ Oe. (a) Real and (b) imaginary parts of $M_{1\omega}(H, T)$ measured as a function of magnetic field at fixed temperatures in the range $T = 129 - 133$ K. (c) Real and (d) imaginary parts of $M_{1\omega}(H,T)$ measured for a frequency of $f = 1111$ Hz.

FIG. 3. Comparison of the magnetic field dependence of the dc differential susceptibility, $dM/dH$, the ac susceptibility, $M'_{1\omega}/h = \chi'_{1\omega}$, at $f = 11$ Hz (right axis), and ac magnetic loss, $M''_{1\omega}/h = \chi''_{1\omega}$, at $f = 11, 111$ Hz (left axis). The dashed lines mark the field regime of the deviation between the dc and ac susceptibilities and the corresponding onset and destruction of ac magnetic loss. The maximum in $dM/dH$ vs $H$, defined as $H_{dM/dH}^{peak}$, is dependent on measurement temperature and occurs at $H \sim 500$ Oe at $T = 129$ K.

FIG. 4. Magnetic field dependence of the linear and nonlinear components of the ac magnetic response for selected temperatures, $M_{n\omega}(H, T)$, at $f = 111$ Hz. The magnitude of all harmonic components are highly dependent on frequency and hence do not display maximum magnitude simultaneously. (a) The ac magnetic loss, $M''_{1\omega}(H)$. The large absolute magnitudes of (b) the third, $M_{3\omega}$, and (c) fifth, $M_{5\omega}$, harmonic response show a remarkably similar field dependence to the ac magnetic loss term. (d) $M'_{1\omega}$. The even harmonics (e) $M_{2\omega}$ and (f) $M_{4\omega}$, have sizeable contributions to the total magnetic response and closely follow the inflection points of $M''_{1\omega}$.

FIG. 5. The third harmonic of the ac magnetic response as a function of magnetic field, $M_{3\omega}(H)$, for $f = 25 - 10,000$ Hz measured at $T = 129$ K.

FIG. 6. Magnetic field dependence of the second harmonic of the ac magnetic response measured with $f = 111$ Hz. (a) $M_{2\omega}(H)$ measured at 130.5 K. Inset: Real and imaginary components of $M_{2\omega}(H)$. (b) Surface plot of $M'_{2\omega}(H, T)$.

FIG. 7. Frequency dependence of the linear ac magnetic response at $T = 129$ K. (a) Real and (b) imaginary parts of the linear ac magnetic response, $M'_{1\omega}(H)$ and $M''_{1\omega}(H)$, measured as a function of magnetic field. In- and out-of-phase linear susceptibility (c) $\chi'_{1\omega} = M'_{1\omega}h$ and (d) $\chi''_{1\omega} = M''_{1\omega}h$ as a function of frequency for the magnetic field corresponding to the dip in $M'_{1\omega}(H)$ and $M''_{1\omega}(H)$ at $H_{dM/dH}^{peak}$ marked by a blue asterisk in (b). The green line represents a fit of (4) and (5) to the low frequency side of the inflection point in $\chi'_{1\omega}$ and the peak in $\chi''_{1\omega}$, respectively, which correspond to $\tau_0 = 1/(2\pi f_0)$. The blue lines represent fits on the high temperature side of $\tau_0$.

FIG. 8. Dynamic parameters as a function of temperature (a) $\tau_0$, the characteristic time, and (b) $\chi''_0$, frequency-independent term, extracted from fits of equation (5) to the low frequency side of $\chi''_{1\omega}$. The corresponding magnetic fields are marked by asterisks in Fig. 7(b).

FIG. 9. The third harmonic of the ac magnetic response as a function of temperature and magnetic field. (a) Frequency dependence of real component of $M_{3\omega}(H)$ for $f = 25 - 10{,}000$ Hz. Inset: $M''_{3\omega}$. (b) – (d) Surface plots of the real part, $M'_{3\omega}(H, T)$, for various frequencies.

FIG. 10. Temperature dependence of the characteristic frequency, $f_0$, corresponding to the magnetic fields marked in Fig. 7(b). Inset: $\chi''_{1\omega}$ vs $f$ for temperatures ranging from $T = 129 - 131.5$ K. The peak in $\chi''_{1\omega}$ at $f_0$ shifts to higher frequency as temperature increases, as indicated by the black arrow, toward the tricritical point.

FIG. 11. Linear and nonlinear magnetic response of the unpolarized CHM state as a function of temperature at $H = 0$ Oe and $f = 10{,}000$ Hz. The kink point at $T_0 \sim 132.25$ K marks the PM-CHM phase transition. Non-zero values of $M_{2\omega} - M_{5\omega}$ appear at $T^* = 133$ K and correspond with the inflection point in $M_{1\omega}$ that marks the onset of chiral correlations in the fluctuation-disordered precursor region.

FIG. 12. Temperature dependence of the ac magnetic response measured with an ac field amplitude, $h = 5$ Oe. (a) Real and (b) imaginary parts of $M_{1\omega}(T)$ measured as a function of temperature with fixed dc fields in the range $H_{dc} = 50 - 1200$ Oe and $f = 111$ Hz. $T_m$ marks the PM–FP transition, which shifts to higher temperature with increasing magnetic field. The inset shows $M_{2\omega}$ which demonstrates the onset of strong FM correlations at $T_m$. (c) $M_{2\omega}$ and (d) $M_{3\omega}$ as a function of temperature for fixed dc fields in the range $H_{dc} = 50 - 400$ Oe at $f = 10{,}000$ Hz.

FIG. 13. Determination of the critical values of field and temperature. (a) First derivative of $M'_{1\omega}$, $M_{3\omega}$, and $M_{2\omega}$ as a function of magnetic field. $H_{C,1}$ is defined where the onset of rapid changes of slope in the linear and nonlinear magnetic response coincide. $H_{C,1}$, represents the crossover field for the onset of the HNL CSL. $H_{C,2}$ defines the critical field for the FFM transition. $dM_{3\omega}/dH$ and $dM_{2\omega}/dH$ are multiplied by a factor of 10. (b) Magnetic field dependence of $M_{1\omega}(T)$ for $H_{dc} = 400 - 550$ Oe measured for $f = 10{,}000$ Hz. As magnetic field decreases, the peak at $T_m$ shifts to lower temperature and disappears at the tricritical point. Inset: Example of Lorentzian peak fit used to determine an accurate value of $T_m$. Standard error obtained from fits of each data set range between $\pm 0.012 - 0.045$ K.

FIG. 14. *H-T* phase diagram determined from the linear and nonlinear components of the ac magnetic response plotted onto (a) $M'_{1\omega}(H,T)$, (b) $M''_{1\omega}(H,T)$, (c) - (d) $M'_{n\omega}(H,T)$ for ($n = 2 - 5$) at $f = 10{,}000$ Hz, where $M_{n\omega}(H,T)$ refers to the field dependence at fixed temperature. In (a) the critical temperatures are labeled: $T^* = 133$ K, $T_0 = 132.3$ K, $T_{\text{TCP}} = 131.35$ K (red dashed line) and $T_{\text{C}} = 130.75$ K (black dashed line). Field-dependent anomalies, marked by open symbols, in $M''_{1\omega}$ (diamonds), $M_{2\omega}$ (circles), and $M_{3\omega}$ (squares) locate the low field increase at $H_{\text{C},1}$ (blue), dip in magnitude at $H_{\text{dM/dH}}^{\text{peak}}$ (blue), and high-field destruction at $H_{\text{C},2}$ (black) of the magnetic loss and higher harmonics corresponding to the HNL CSL. Equivalent temperature-dependent anomalies from $M_{n\omega}(T,H_{\text{dc}})$ measurements are marked by closed symbols. The IC-C phase line is given by the fall off of $M'_{1\omega}(H)$ (stars). Above $T_{\text{TCP}}$, the red symbols mark the lower and upper bounds of the anomaly across the linear CSL-PM phase transition. The magnetic loss and nonlinear response abruptly fall off at $T_{\text{TCP}} = 131.35$ K given by the intersection of the PM-FP line defined by $T_{\text{m}}$ (green/pink squares) and the IC-C phase line determined from the power law fit of $H_{\text{C},2}$ (solid black line).

FIG. 1

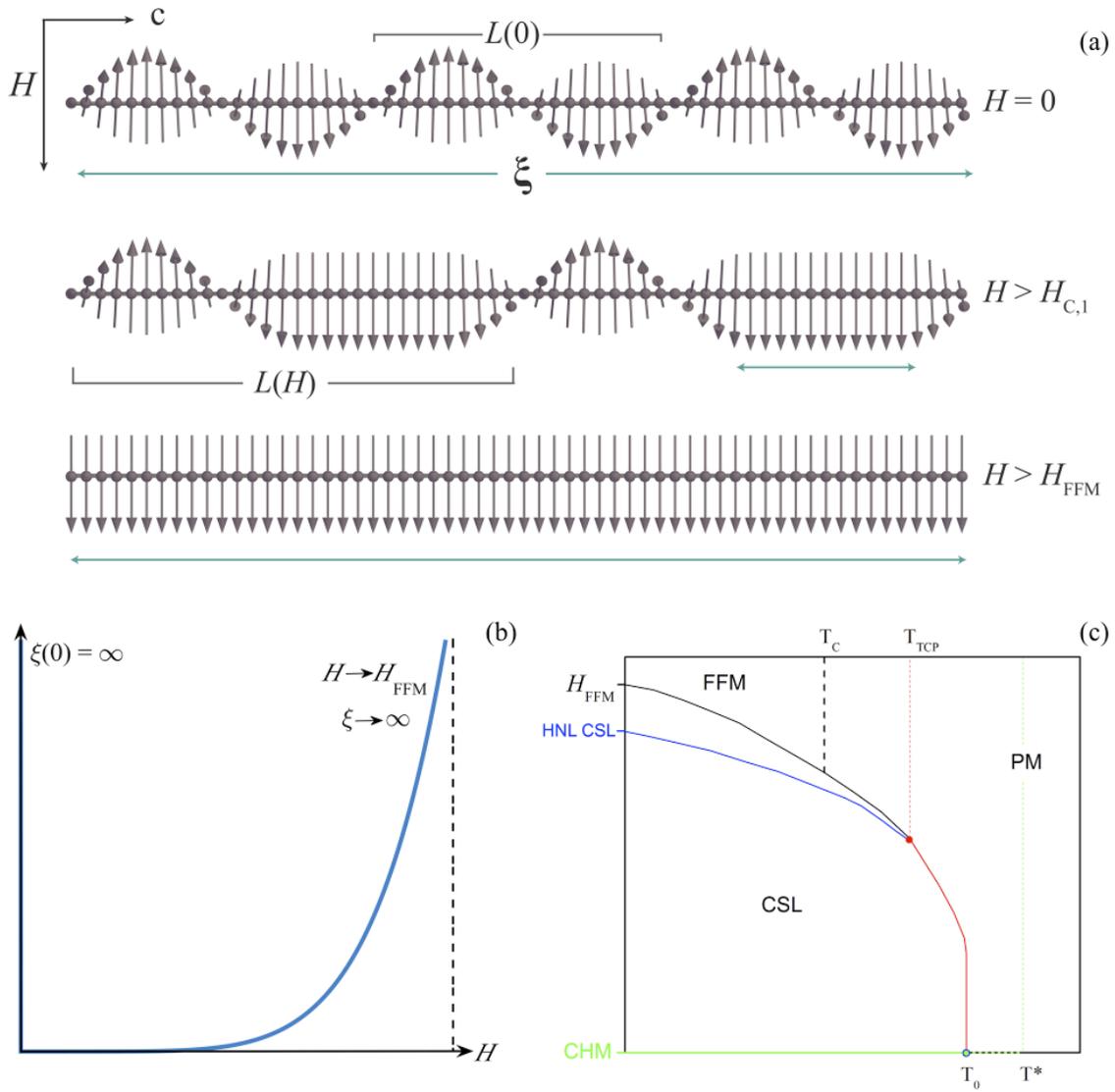

FIG. 2

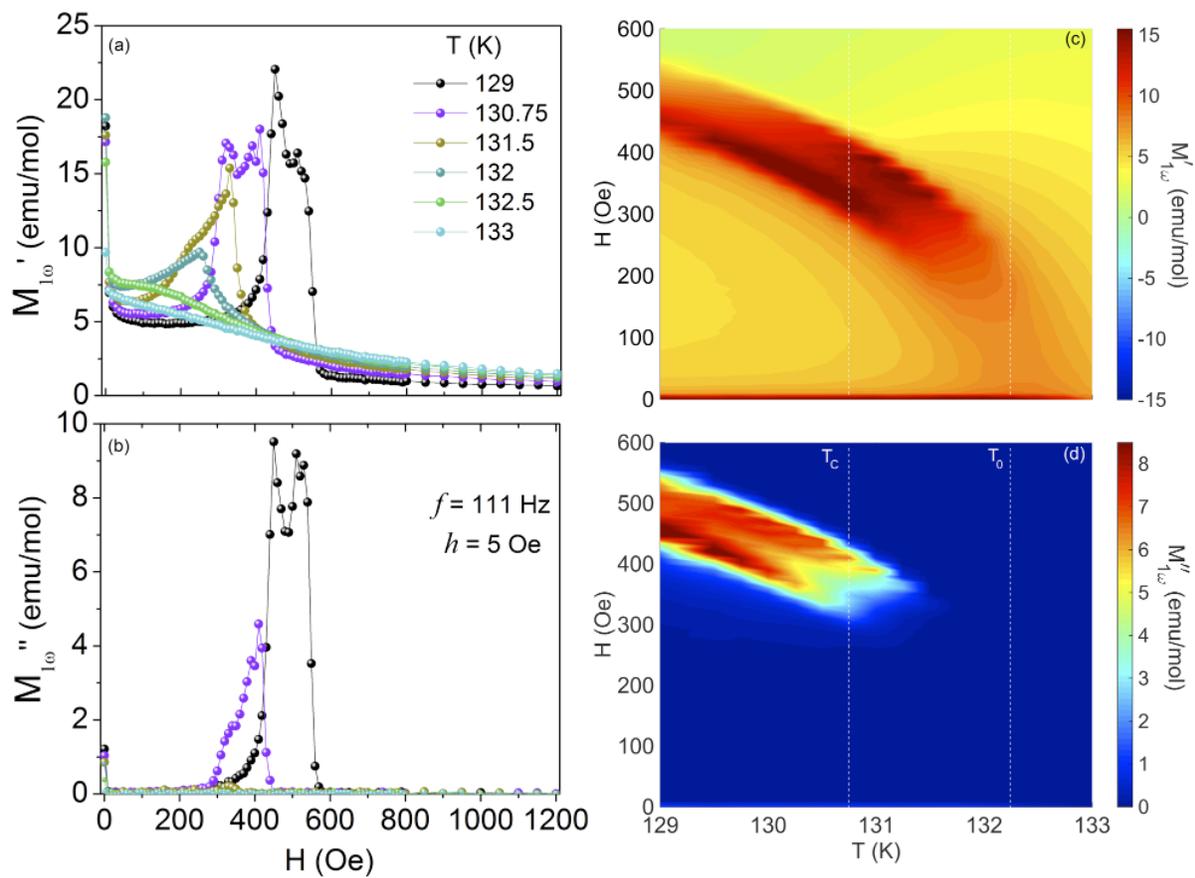

FIG. 3

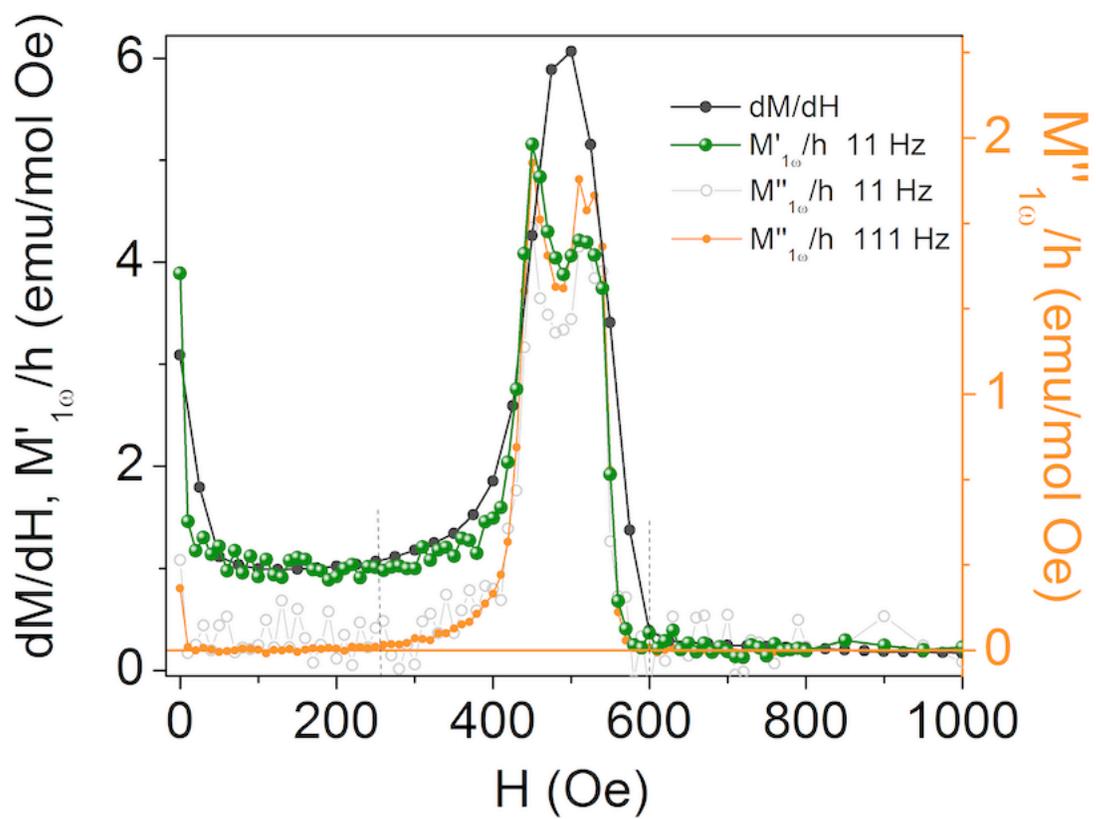

FIG. 4

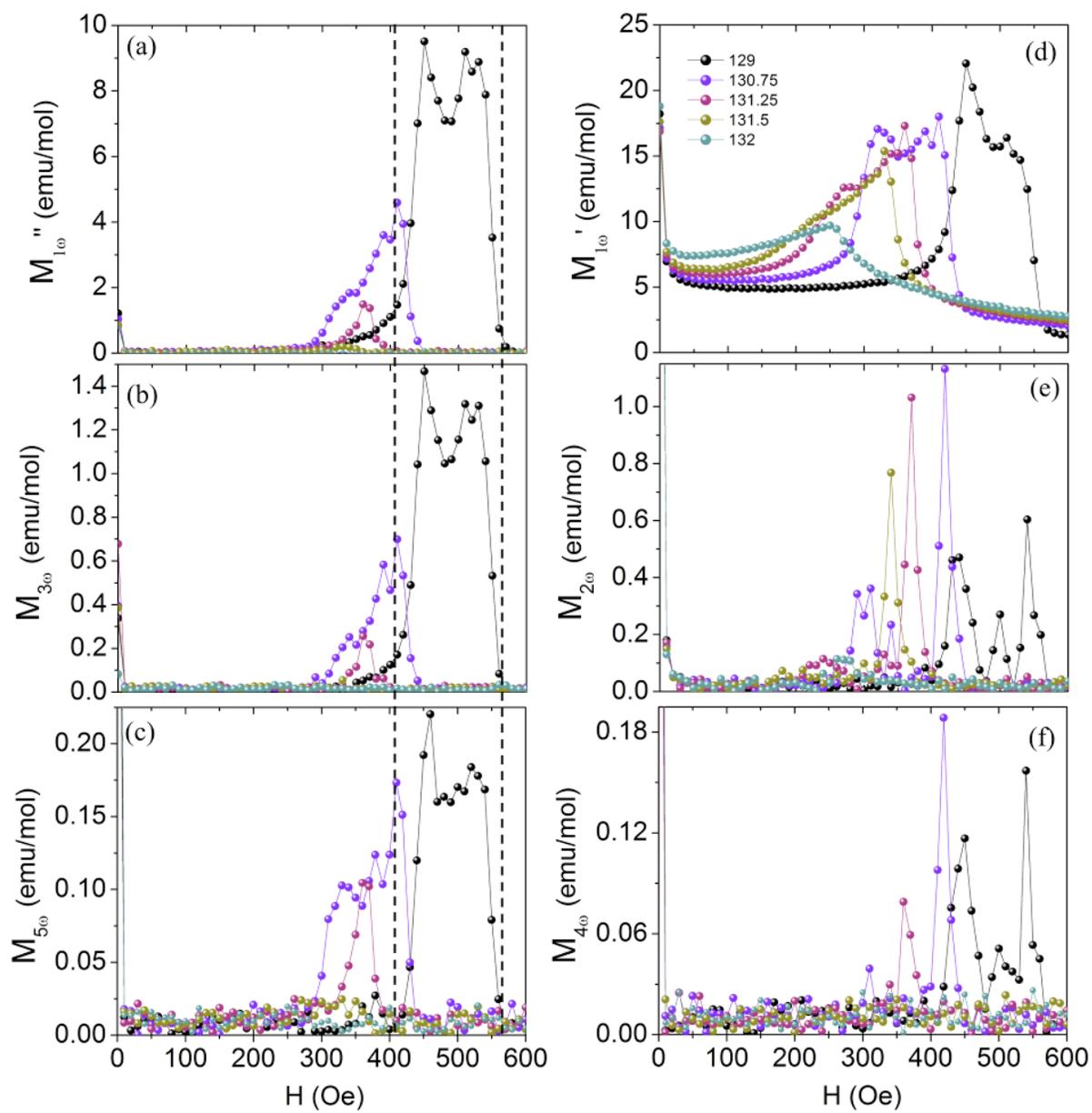

FIG. 5

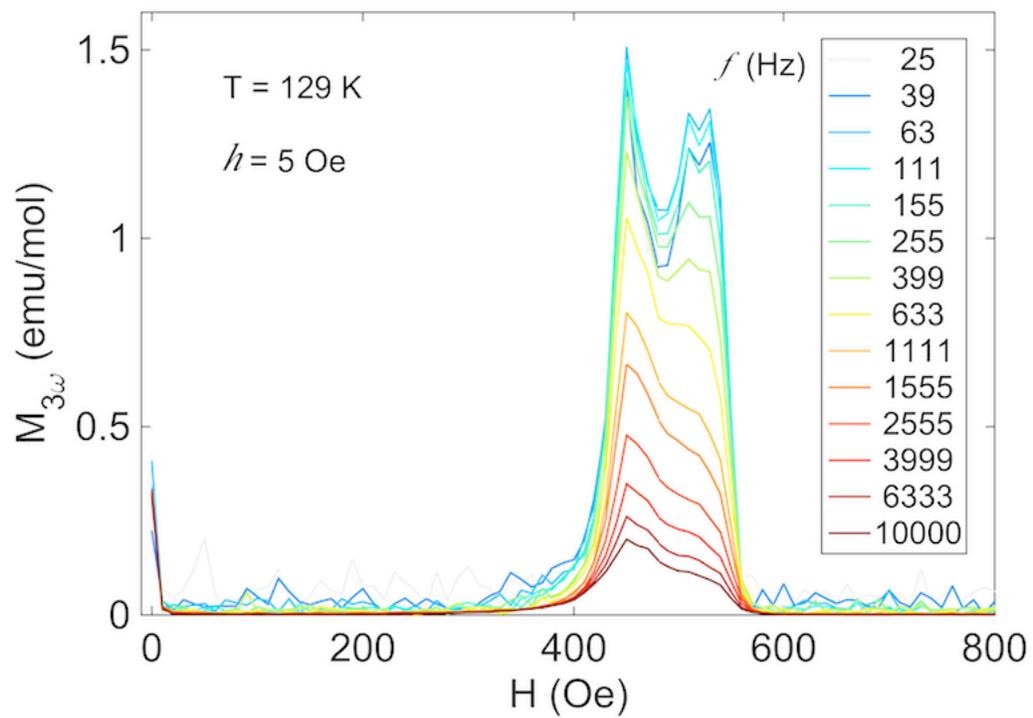

FIG. 6

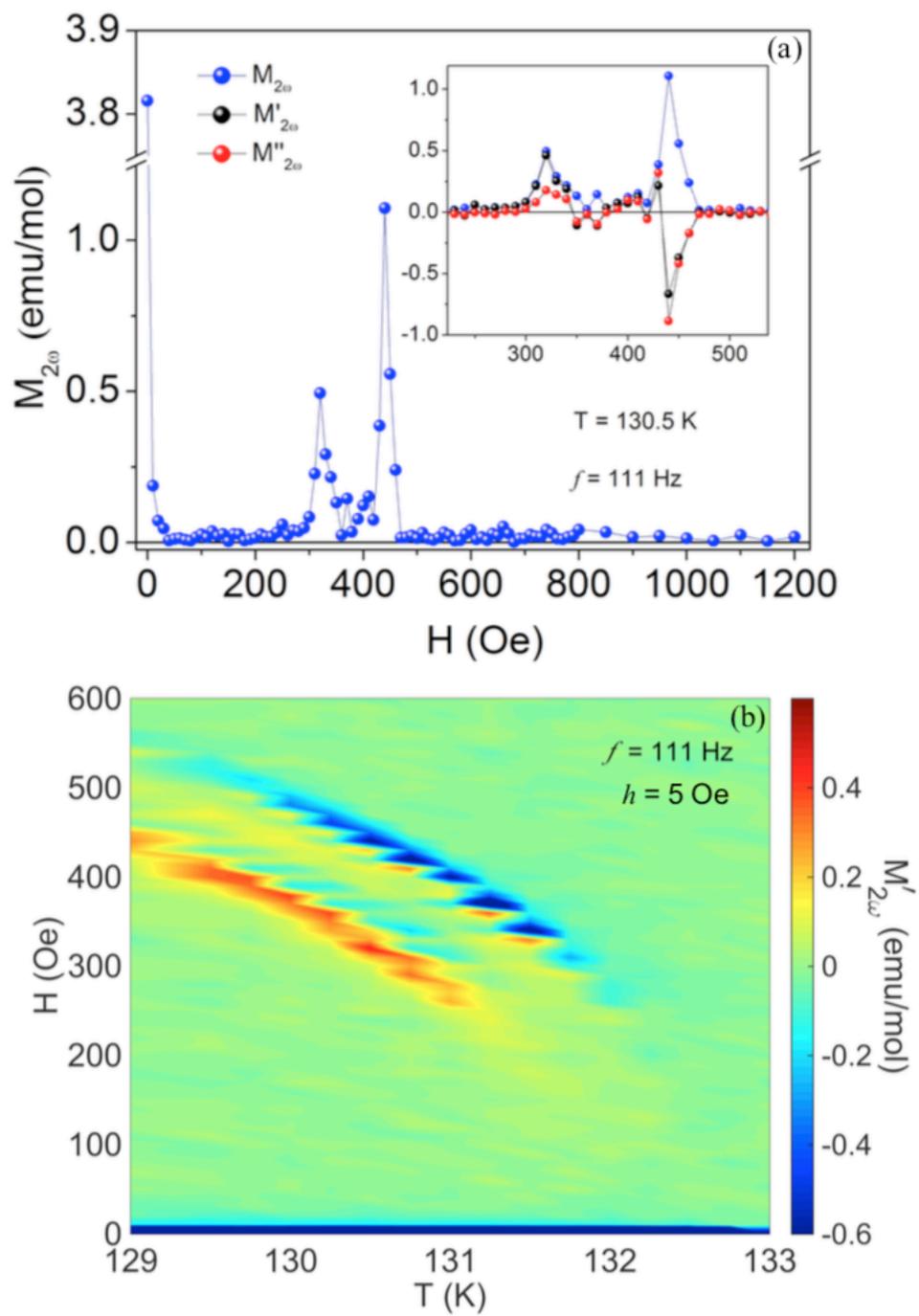

FIG. 7

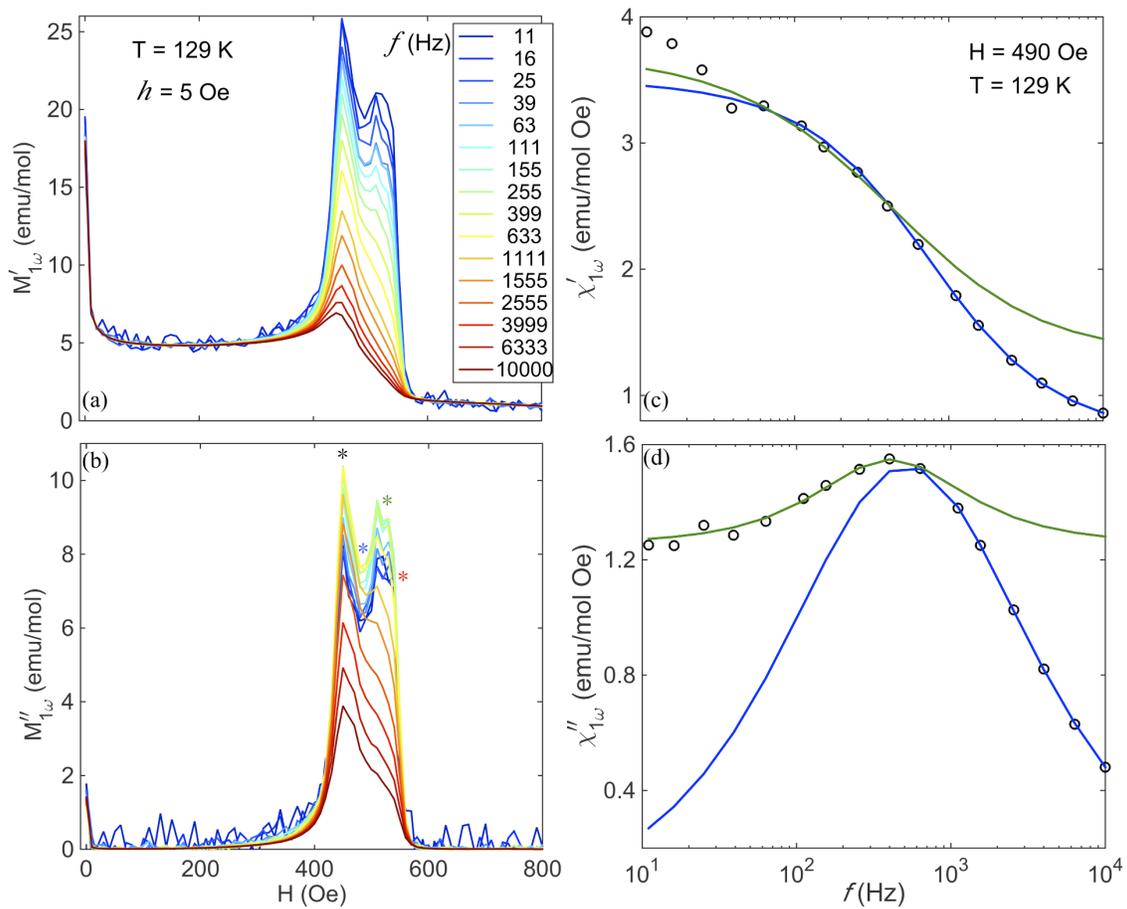

FIG. 8

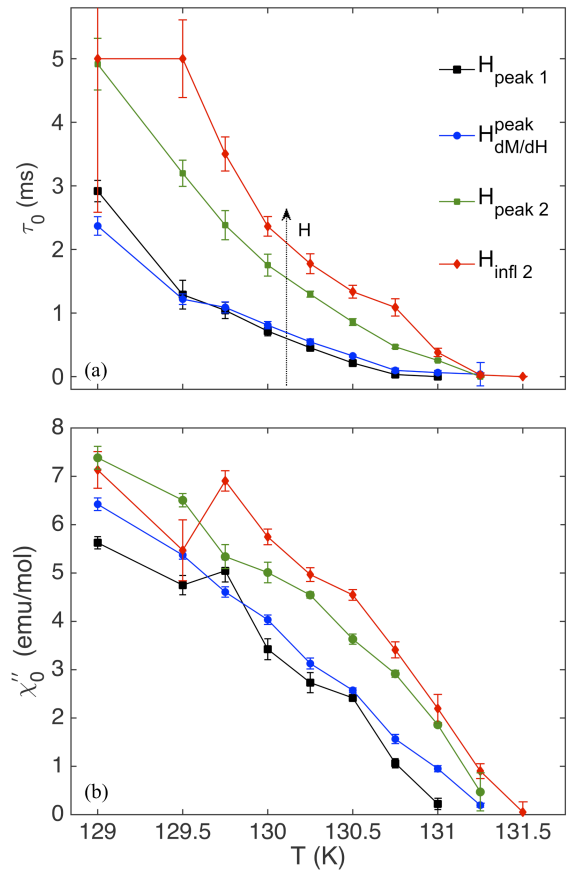

FIG. 9

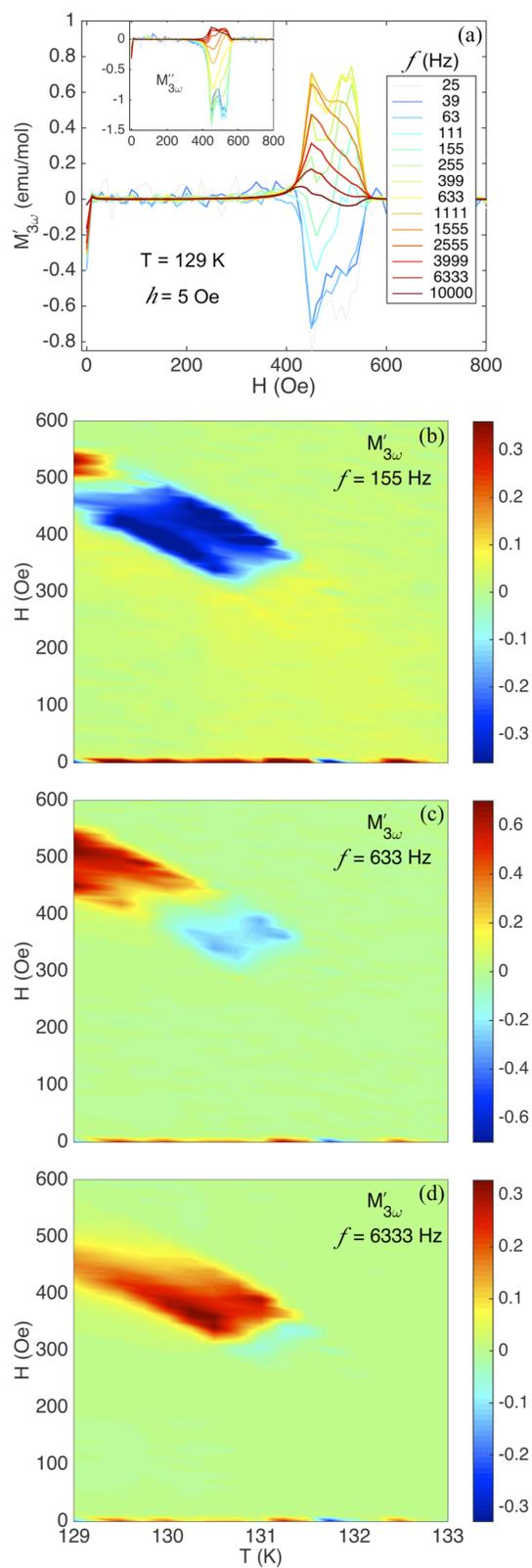

FIG. 10

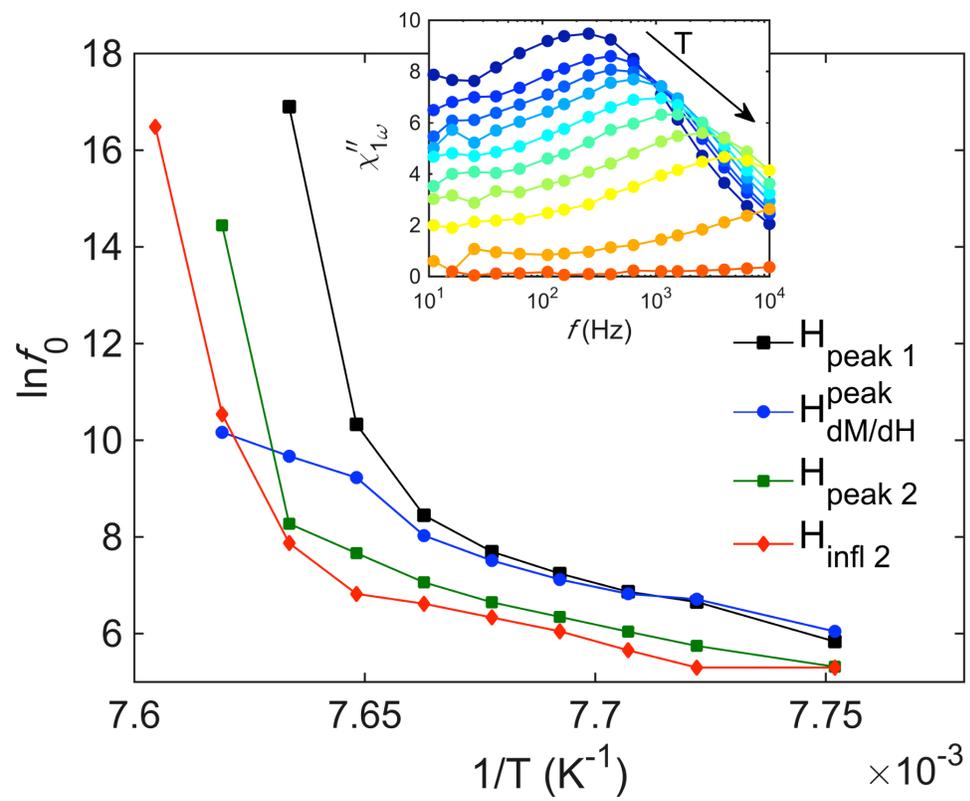

FIG. 11

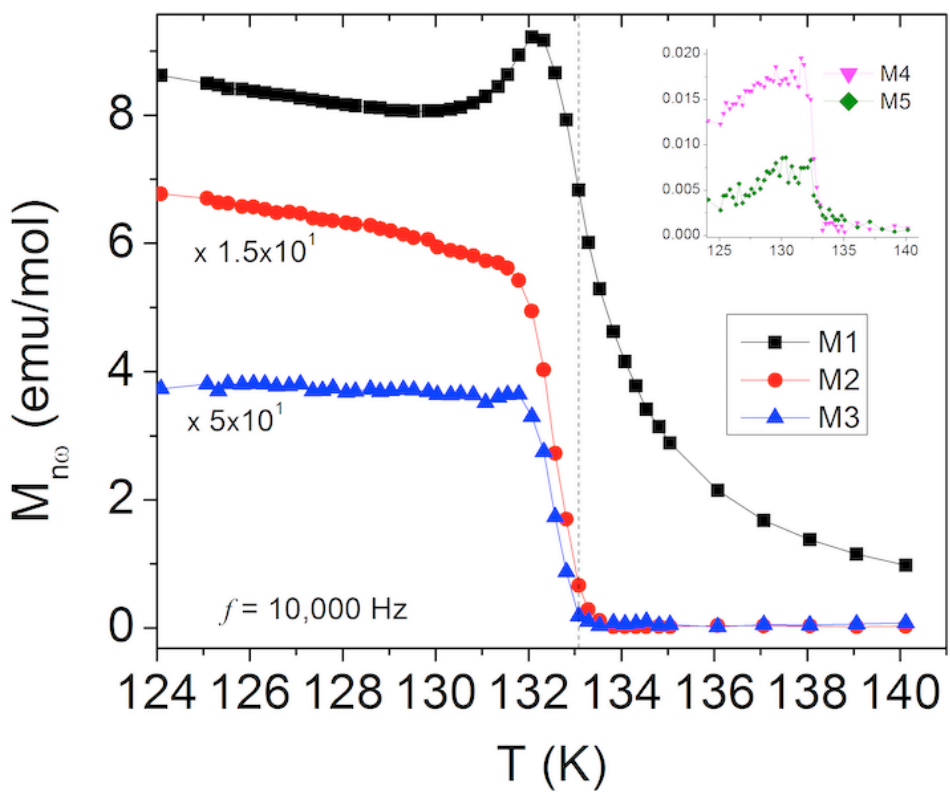

FIG. 12

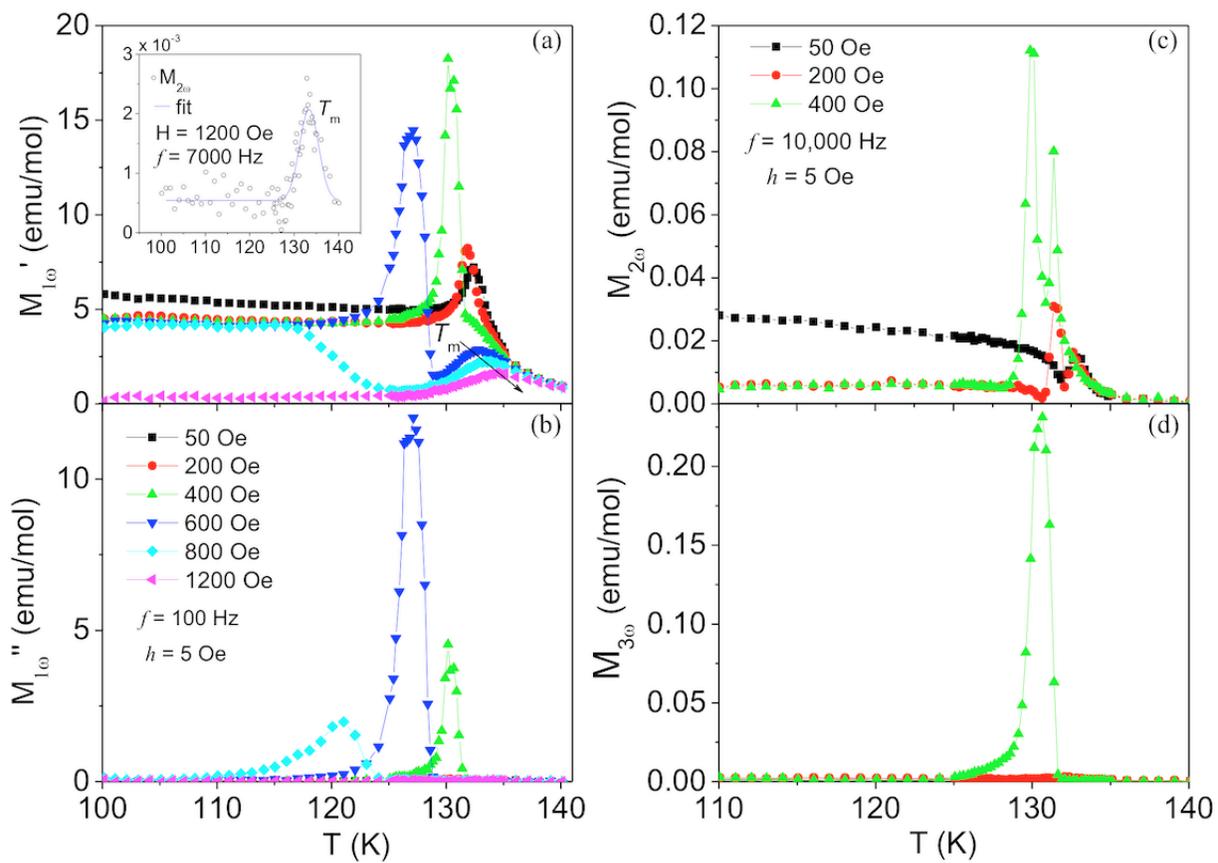

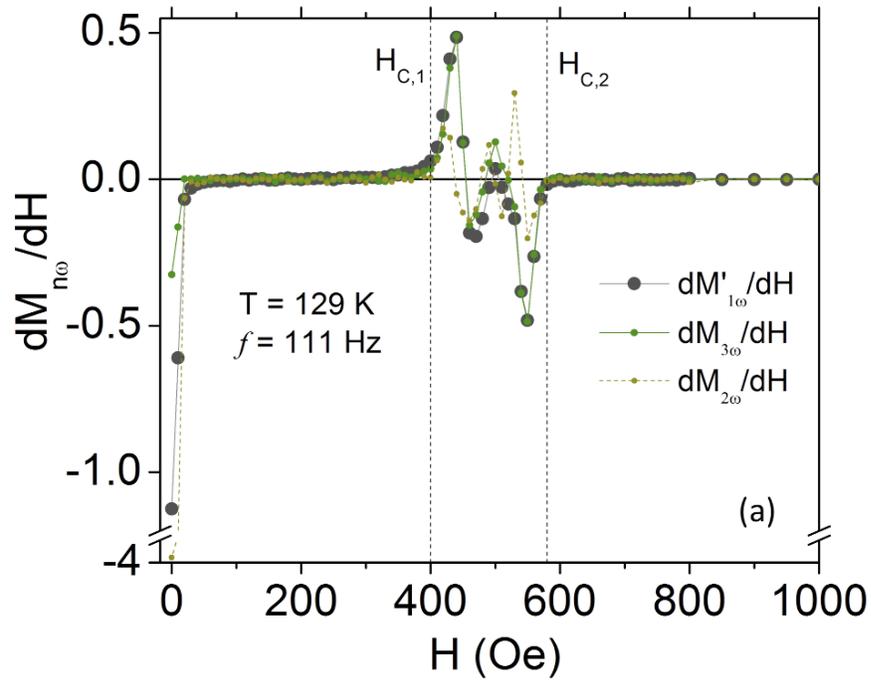

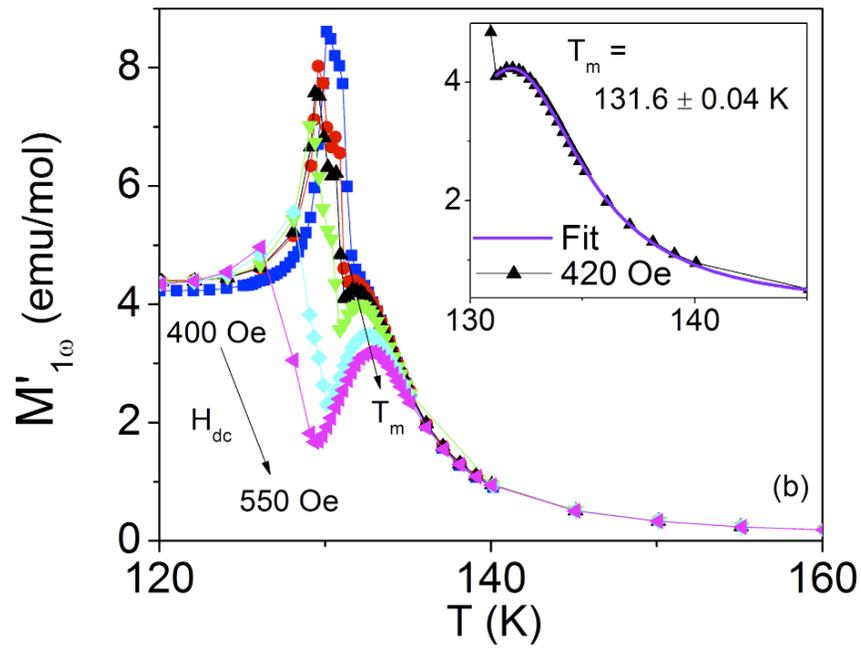

FIG. 14

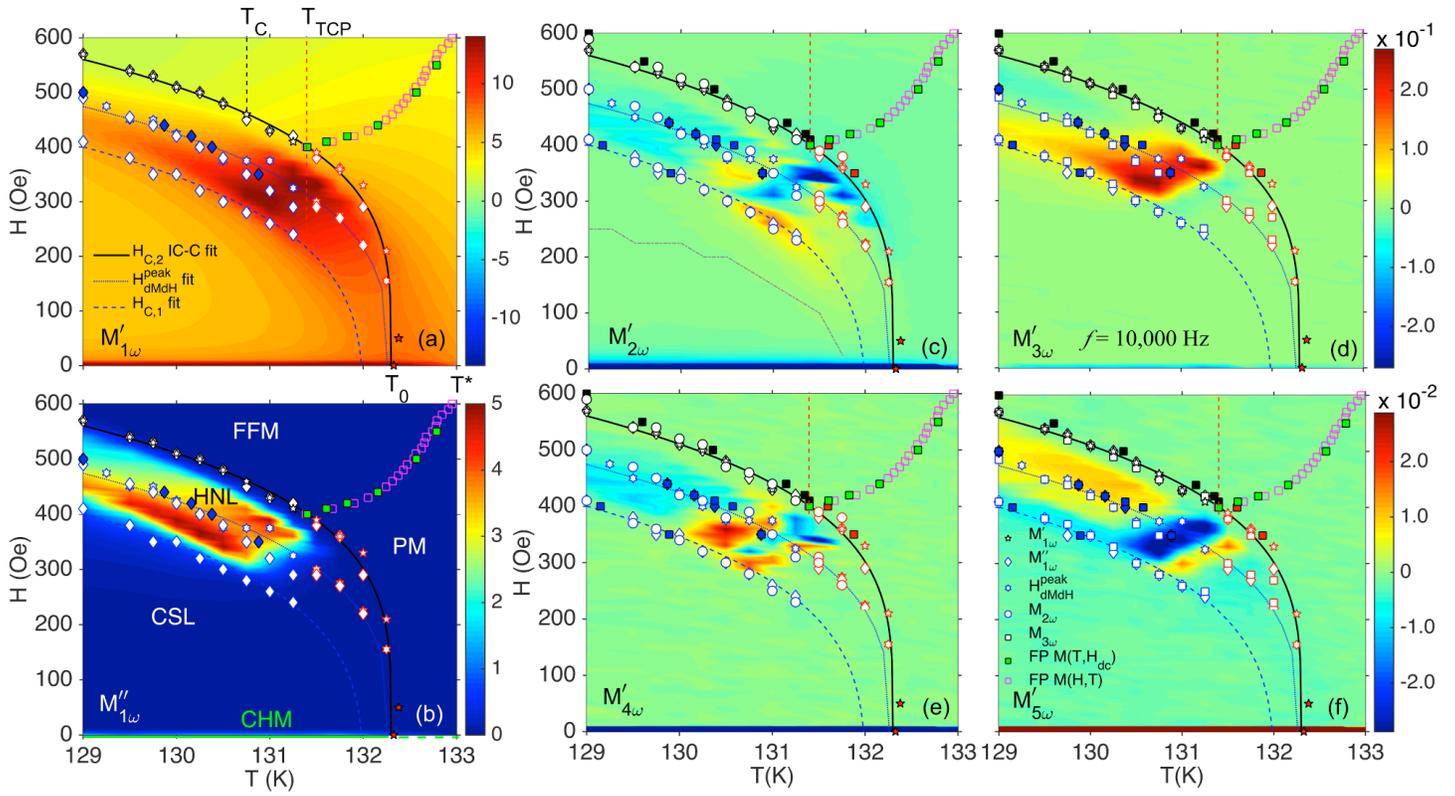